\documentclass[a4paper]{spie}  % use this instead for A4 paper
\usepackage{amsmath,amsfonts,amssymb}

\usepackage[utf8]{inputenc}   % lettere accentate
\usepackage[english]{babel}   % sillabazione
\usepackage[usenames,dvipsnames]{xcolor} % for hyperref
\usepackage[colorlinks=true, allcolors=blue]{hyperref}
\usepackage{aas_macros}
\usepackage{paralist}
\usepackage{graphicx}

\title{Development status of the SOXS instrument control software}

\author[*,a]{Davide~Ricci}
\author[a]{Andrea~Baruffolo}
\author[a]{Bernardo~Salasnich}
\author[a]{Marco~De~Pascale}

\author[b]{Sergio~Campana}
\author[a]{Riccardo~Claudi}
\author[c]{Pietro~Schipani}
\author[b]{Matteo~Aliverti}
\author[d]{Sagi~Ben-Ami}
\author[a,i]{Federico~Biondi}
\author[c]{Giulio~Capasso}
\author[e]{Rosario~Cosentino}
\author[f]{Francesco~D'Alessio}
\author[b]{Paolo~D'Avanzo}
\author[g]{Ofir	Hershko}
\author[j,q]{Hanindyo~Kuncarayakti}
\author[b]{Marco~Landoni}
\author[k]{Matteo~Munari}
\author[m,t]{Giuliano~Pignata}
\author[a]{Kalyan~Radhakrishnan}
\author[h]{Adam~Rubin}
\author[k]{Salvatore~Scuderi}
\author[f]{Fabrizio~Vitali}
\author[s]{David~Young}
\author[l]{Jani~Achrén}
\author[m,t]{José~Antonio~Araiza-Duran}
\author[n]{Iair~Arcavi}
\author[h]{Anna~Brucalassi}
\author[g]{Rachel~Bruch}
\author[a]{Enrico~Cappellaro}
\author[c]{Mirko~Colapietro}
\author[c]{Massimo~Della~Valle}
\author[k]{Rosario~Di~Benedetto}
\author[c]{Sergio~D'Orsi}
\author[g]{Avishay~Gal-Yam}
\author[b]{Matteo~Genoni}
\author[e]{Marcos~Hernandez}
\author[j,q]{Jari~Kotilainen}
\author[r]{Gianluca~Li~Causi}
\author[q]{Seppo~Mattila}
\author[s]{Michael~Rappaport}
\author[b]{Marco~Riva}
\author[s]{Stephen~Smartt}
\author[k]{Ricardo~Zanmar~Sanchez}
\author[u]{Maximilian~Stritzinger}
\author[e]{Hector~Ventura}

\affil[a]{INAF -- Osservatorio Astronomico di Padova, Vicolo dell’Osservatorio 5, I-35122, Padua, Italy }
\affil[b]{INAF -- Osservatorio Astronomico di Brera, Via Bianchi 46, I-23807, Merate, Italy }
\affil[c]{INAF -- Osservatorio Astronomico di Capodimonte, Sal. Moiariello 16, I-80131, Naples, Italy }
\affil[d]{Harvard-Smithsonian Center for Astrophysics, Cambridge, USA }
\affil[e]{FGG-INAF, TNG, Rambla J.A. Fernández Pérez 7, E-38712 Breña Baja (TF), Spain }
\affil[f]{INAF -- Osservatorio Astronomico di Roma, Via Frascati 33, I-00078 M. Porzio Catone, Italy }
\affil[g]{Weizmann Institute of Science, Herzl St 234, Rehovot, 7610001, Israel }
\affil[h]{ESO, Karl Schwarzschild Strasse 2, D-85748, Garching bei München, Germany }
\affil[i]{Max-Planck-Institut für Extraterrestrische Physik, Giessenbachstr. 1, D-85748 Garching, Germany }
\affil[j]{Finnish Centre for Astronomy with ESO (FINCA), FI-20014 University of Turku, Finland}
\affil[k]{INAF -- Osservatorio Astrofisico di Catania, Via S. Sofia 78 30, I-95123 Catania, Italy }
\affil[l]{Incident Angle Oy, Capsiankatu 4 A 29, FI-20320 Turku, Finland }
\affil[m]{Universidad Andres Bello, Avda. Republica 252, Santiago, Chile }
\affil[n]{Tel Aviv University, Department of Astrophysics, 69978 Tel Aviv, Israel }
\affil[o]{Dark Cosmology Centre, Juliane Maries Vej 30, DK-2100 Copenhagen, Denmark }
\affil[p]{Aboa Space Research Oy, Tierankatu 4B, FI-20520 Turku, Finland}
\affil[q]{Tuorla Observatory, Dept. of Physics and Astronomy, FI-20014 University of Turku, Finland }
\affil[r]{INAF - Istituto di Astrofisica e Planetologia Spaziali, Rome, Italy}
\affil[s]{Astrophysics Research Centre, Queen's University Belfast, Belfast, BT7 1NN, UK }
\affil[t]{Millennium Institute of Astrophysics (MAS)}
\affil[u]{Aarhus University, Ny Munkegade 120, D-8000 Aarhus, Denmark }

\authorinfo{$^*$Contact information:
  D.R: davide.ricci@inaf.it, +39-049-829-3480
}

\begin{document}
\maketitle

\begin{abstract}
  SOXS (Son Of X-Shooter) is a forthcoming instrument for ESO-NTT,
  mainly dedicated to the spectroscopic study of transient events and
  is currently starting the AIT (Assembly, Integration, and Test)
  phase.  It foresees a visible spectrograph, a near-Infrared (NIR)
  spectrograph, and an acquisition camera for light imaging and
  secondary guiding.
  The optimal setup and the monitoring of SOXS are carried out with a
  set of software-controlled motorized components and sensors. The
  instrument control software (INS) also manages the observation and
  calibration procedures, as well as maintenance and self-test
  operations.
  %SOXS has successfully passed the FDR, and most of
  %optomechanics are in Padova, ready for the AIT phase.
  % 
  The architecture of INS, based on the latest release of the VLT
  Software (VLT2019), has been frozen; the code development is in an
  advanced state for what concerns supported components and
  observation procedures, which run in simulation.

  In this proceeding we present the INS current status, focusing in
  particular on the ongoing efforts in the support of two
  non-standard, ``special'' devices.
  The first special device is the piezoelectric slit exchanger for the
  NIR spectrograph;
  the second special device is the piezoelectric tip-tilt corrector
  used for active compensation of mechanical flexures of the
  instrument.
  For both, which are commanded via a serial line, specific driver and
  simulators have been implemented.
  %
  % The third device is the COTS camera for light imaging, interfaced
  % to INS through the Technical Detector Control Software component of
  % the VLTSW.  The special feature is the secondary guiding procedure,
  % which allows query, selection and trying of star candidates when the
  % camera is used for automatic star tracking.
\end{abstract}


%%%%%%%%%%%%%%%%%%%%%%%%%%%%%%%%%%%%%%%%%%%%%%%%%%%%%%%%%%%%%%%%
\section{Introduction}
\label{sec:intro}

The SOXS instrument \textit{``Son Of X-Shooter''}, a forthcoming
facility\cite{2016SPIE.9908E..41S} for the European Southern
Observatory (ESO) New Technologies Telescope (NTT) telescope at the La
Silla Observatory, Chile, successfully passed the Final Design Review
(FDR) process on July 2018, and it is approaching the Assembly,
Integration and Test (AIT) phase of its several subsystems\cite{
  2018SPIE10707E..2HC, 2018SPIE10707E..1GR, 2018SPIE10702E..3TC,
  2018SPIE10702E..3DB, 2018SPIE10702E..31A, 2018SPIE10702E..2ZR,
  2018SPIE10702E..2MB, 2018SPIE10702E..2JC, 2018SPIE10702E..28V,
  2018SPIE10702E..27Z, 2018SPIE10702E..0FS, 2018SPIE10702E..04P}:
\begin{inparaenum}
\item The Common Path (CP);
\item The Visible spectrograph (UV-VIS);
\item The Near-Infrared spectrograph (NIR);
\item The Acquisition Camera (AC);
\item The Calibration Unit (CU).
\end{inparaenum}

% soxsscuderi

This paper is part of a series of contributions \cite{soxsschipani,
  soxslandoni, soxsrubin, soxsbiondi, soxsgenoni,
  soxskuncarayakti, soxsyoung, soxsbrucalassi, soxscolapietro,
  soxscosentino, soxsclaudi, soxsaliverti, soxssanchez, soxsvitali}
describing the current development status of the SOXS subsystems.
In particular, we present the progresses in the status of the
Instrument control Software (INS) following the last dedicated
proceedings\cite{2018SPIE10707E..1GR} and we focus on two special
devices which required a custom development: the Near Infrared Slit
Exchanger (NISE) and the Active Flexure Compensator (AFC).

The control network architecture and software design architecture are
presented in Sect.~\ref{sec:net}.
The development of the NISE is shown in Sect.~\ref{sec:nise}, while the
development of the AFC is treated in Sect.~\ref{sec:afc}.
Conclusions are presented in Sect.~\ref{sec:conc}.

%%%%%%%%%%%%%%%%%%%%%%%%%%%%%%%%%%%%%%%%%%%%%%%%%%%%%%%%%%%%%%%%
\section{Network and Software architecture}
\label{sec:net}

%% -------------------------------------------------------------
\begin{figure} [t]
  \centering
  \begin{tabular}{cc}
    \includegraphics[width=0.452\textwidth]{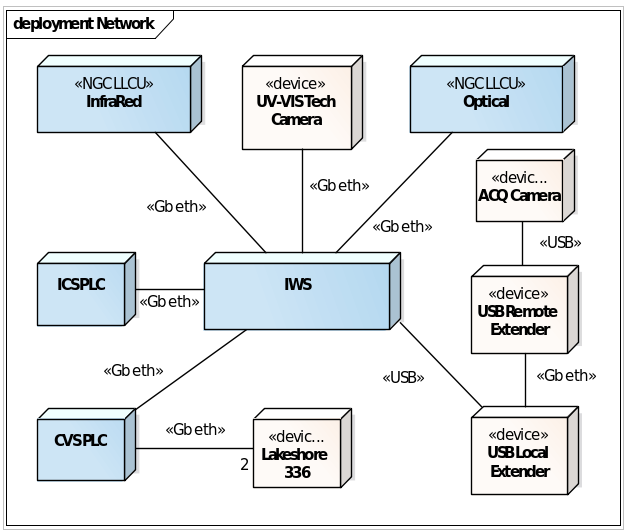}
    \includegraphics[width=0.510\textwidth]{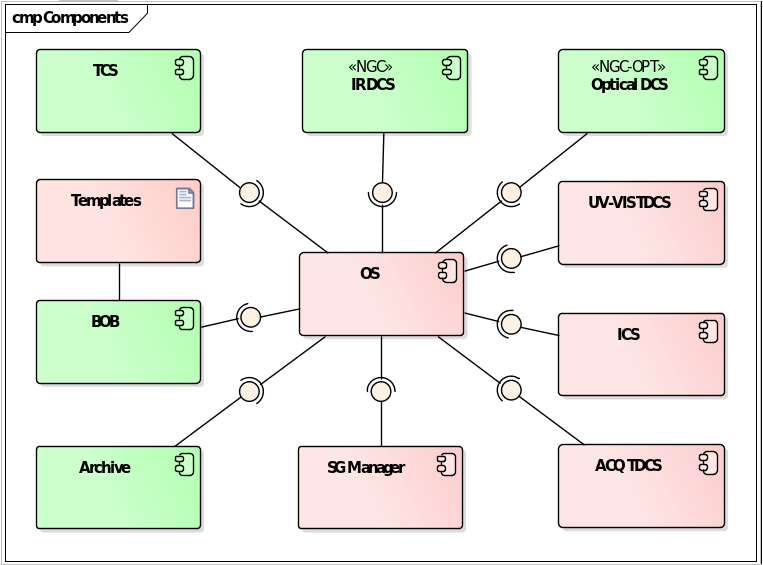}
  \end{tabular}
  \caption[control-network]
  { \label{fig:control-network} Left: Network control architecture of
    SOXS.  Right: Components of the SOXS software; red boxes represent
    software requiring custom configuration or development,
    green boxes represent VLTSW components that will be used without
    modifications.}
\end{figure}
%% -------------------------------------------------------------

The SOXS network architecture follows the typical configuration of
VLT Instruments control systems: an Instrument Workstation (IWS)
supervises through the instrument LAN several connected local
controllers, mostly based based on Gb Ethernet (see
Fig.~\ref{fig:control-network}~left).

In particular, for SOXS, two ESO New General Detector Controllers
(NGC) are responsible of the UV-VIS and NIR detectors, while the
commercial AC camera, providing an integrated controller with USB
interface, is linked to the IWS through a commercial ICRON USB
extender.  An additional Cameralink Technical Camera (TECH),
physically placed in the UV-VIS spectrograph subsystem, is also linked
via Gb Ethernet.

A single \emph{Beckhoff} Programmable Logic Controller
(PLC)\cite{2014SPIE.9152E..07K} is responsible for the control of all
instrument functions, while a separate \emph{Siemens} S7 PLC
autonomously controls Cryo-Vacuum functions and the privately
Ethernet-connected \emph{Lakeshore} 336 temperature controller.

%%%%%%%%%%%%%%%%%%%%%%%%%%%%%%%%%%%%%%%%%%%%%%%%%%%%%%%%%%%%%%%%

The SOXS INS (see Fig.~\ref{fig:control-network}~right) is developed
using the latest VLT Software release (VLT2019).
It is in charge of the control of:
\begin{inparaenum}
\item all instrument functions (ICS);
\item the UV-VIS and NIR spectrograph detectors, controlled by instances
  of Detector Control Software (DCS);
\item the AC and the TECH cameras, basing on instances of the
  Technical DCS Software Development Kit (SDK)
  \cite{2014SPIE.9152E..0ID};
\item the observation procedures via the Observation Software (OS),
  managing observation, calibration and maintenance procedures
  implemented as templates and executed by the Broker of Observation
  Blocks (BOB);
\item the external interfaces such as the Telescope Control Software
  (TCS), and the Archive.
\end{inparaenum}
Currently, all these components have been configured and developed, as
well as control panels for and user interfaces, and run in simulation
under the VLT Software environment.

The most of the ICS SOXS components are natively supported as
``standard devices'' and it is sufficient to provide configuration
information.
For non-standard devices, it is required to properly interface them
with ICS, developing a Function Block (FB) software at PLC level and a
``special device'' driver at IWS level.

In SOXS, these special devices are the cryogenic piezo-mechanic stage
for slit positioning in the NIR spectrograph and the piezo-actuated
tip-tilt mirrors used for Active Flexure Compensation.  Details of the
development of these two special devices are given in following
sections.

%%%%%%%%%%%%%%%%%%%%%%%%%%%%%%%%%%%%%%%%%%%%%%%%%%%%%%%%%%%%%%%%
\section{Near Infrared Slit Exchanger}
\label{sec:nise}

%% -------------------------------------------------------------
\begin{figure} [t]
  \centering
  \begin{tabular}{cc}
    \includegraphics[width=0.475\textwidth]{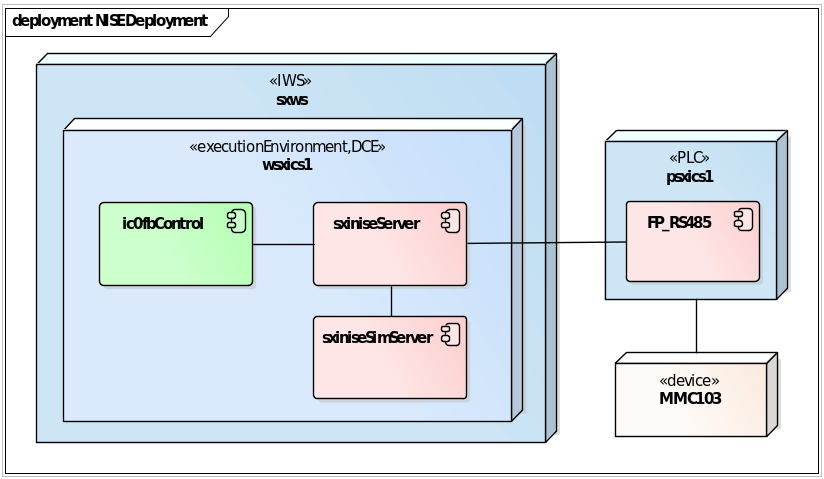}
    \includegraphics[width=0.500\textwidth]{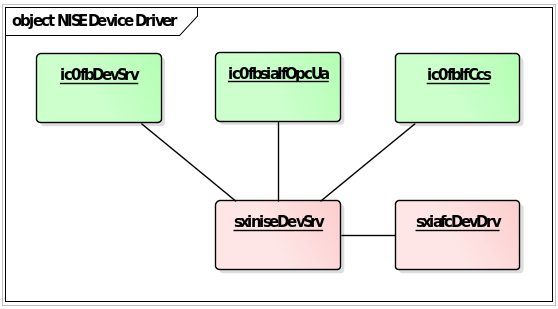}
  \end{tabular}
  \caption[NISE]
  { \label{fig:nise}
    Left: NISE special device deployment diagram.
    Right: NISE Device Driver object diagram.}
\end{figure}
%% -------------------------------------------------------------

The NIR Infrared Slit Exchanger (NISE) is a cryogenic actuator
controlled via a Micronix \texttt{MMC-103} crontroller, connected to
the SOXS PLC through a serial line of type \texttt{RS485}.  Since the
Micronix controller is not directly supported by the VLT Software, a
special device needs to be developed.  The design of the NISE is shown
in Fig.~\ref{fig:nise}~left.
A dedicated device driver class named \texttt{sxiniseDevDrv}, shown in
Fig.~\ref{fig:nise}~right, was derived from
\texttt{ic0fbDevDrvBase}. Methods were developed to implement the
device specific behavior.
State change handling methods handles setting up of the communication
with the controller.  The setup handling method is overloaded to
transform setup requests into commands for the Micronix controller.
The status handling method is overloaded to retrieve status
information from the Micronix controller, returning it as a command
reply and storing it in the database in order to be displayed in GUIs.
As shown in Fig.~\ref{fig:nise}~right, the device server
\texttt{sxiniseDevSrv}, i.e. the process that hosts the driver code,
is based on the standard server class \texttt{ic0fbDevSrv} and makes
use of standard communication interfaces \texttt{ic0fbsiaOpcUa} and
\texttt{ic0fbIfCcs}, in order to communicate with the driver or the
simulator.

%%%%%%%%%%%%%%%%%%%%%%%%%%%%%%%%%%%%%%%%%%%%%%%%%%%%%%%%%%%%%%%%
\section{Active Flexure Compensation}
\label{sec:afc}

% %% -------------------------------------------------------------
% \begin{figure} [t]
%   \centering
%   \begin{tabular}{cc}
%     \includegraphics[width=0.482\textwidth]{afc1}
%     \includegraphics[width=0.500\textwidth]{afc2}
%   \end{tabular}
%   \caption[AFC]
%   { \label{fig:afc}
%     Left: AFC special device deployment diagram.
%     Right: AFC Device Driver object diagram.}
% \end{figure}
% %% -------------------------------------------------------------

Since SOXS will be installed at the Nasmyth focus of the NTT, during
an observation it will change its orientation with respect to the
gravity vector.  This will result in some flexures which might move
the target with respect to the spectrographs slit.  For this reason,
two piezo-actuated tip-tilt mirrors (TTM) are located in the common
path and will be used to correct for this effect.

The TTMs will be commanded by INS through the instrument PLC via
analog signals (one per axis). Since the TTMs are not a VLT standard
actuator, a ``special device'' has been developed.  During
observations, this component will operate as a ``tracking axis'',
updating in a loop the position of the TTM depending on the rotator
angle.

% Since the expected loop frequency is about 1~Hz, timing
% constraints are not tight, so it has been decided to implement this
% tracking loop entirely in the IWS.

These TTMs, placed in the Common Path, will assure Active Flexure
Compensation (AFC) of the UV-VIS (AFC1) and the NIR (AFC2) arm,
respectively.  They are controlled by two \texttt{PI S-330}
two-axis actuators. Each actuator is controlled by a \texttt{PI
  E-727.3SDA} 3 channel digital piezo controller, which is
commanded through the instrument PLC via serial line.  The active
flexure compensation system operates in the following modes:
\begin{inparaenum}
\item Mode AUTO, in which the correction is periodically computed and
  applied (about every minute) by the software on the basis of a
  ``pointing model''. The pointing model requires a calibration
  procedure and the computation of corrections requires information
  about the rotator position. The TTM in the visible arm will also
  correct for ADC ``wobbling'' (if necessary), so will also take the
  ADC prism angle in input.
\item Mode STAT, in which the TTM is kept at a fixed position, sent
  via a SETUP command.
\item Mode REF, which puts the TTM at a fixed, pre-defined, position
  required for the alignment of the system.
\end{inparaenum}

The design of the AFC special device is similar to the one of the NISE
Fig.~\ref{fig:nise}.
A dedicated device driver class, in this case \texttt{sxiafcDevDrv},
is derived from \texttt{ic0fbDevDrvBase}, methods will be developed to
implement the device specific behavior. In particular, method
\texttt{controlLoopUser} encapsulates the logic for TTM
positioning. The method is called periodically by the underlying ICS
framework code.
If the AFC has been setup with a fixed position (either specified by
the user or the reference one), the (fixed) positioning command is
``refreshed''. If the AFC must compensate for flexures, a new TTM
command is computed for the current position of the de-rotator and
applied.  The loop period can be set in the device configuration.

In the case of the AFC, commands are sent to the TTM via serial line.
On the PLC side we developed a function block which uses the library
\texttt{FB\_RS232} provided by ESO to handle the serial connection.  A
device simulator (\texttt{sxiafcDevSim}) allows to operate the SW

% As shown in Right~Fig.~\ref{fig:afc}, the device server
% \texttt{sxiafcDevSrv}, i.e. the process that will host the driver
% code, will be based on the standard server class
% \texttt{ic0fbDevSrv}, and make use of standard communication
% interfaces (\texttt{ic0fbsiaOpcUa} and \texttt{ic0fbIfCcs}) to
% communicate with the driver or the simulator.

%%%%%%%%%%%%%%%%%%%%%%%%%%%%%%%%%%%%%%%%%%%%%%%%%%%%%%%%%%%%%%%% 
\section{Conclusion}
\label{sec:conc}

We presented the progresses in the development of the Instrument
Control Software of the forthcoming SOXS instrument, based on the VLT
Software.
We focused on the software development of the two non standard
devices: the Near Infrared Slit Exchanger and the Active Flexure
Compensation system.
Further configuration, development and tests are ongoing in order to
complete the AIT phase of the several subsystems, which is starting in
these months.

% \acknowledgments

% References
\bibliographystyle{spiebib} % makes bibtex use spiebib.bst
\bibliography{ricci-soxs} % bibliography data in ricci-soxs.bib

\end{document}